\newcommand{\beq}{\begin{equation}}
\newcommand{\eeq}{\end{equation}}
\newcommand{\beqa}{\begin{eqnarray}}
\newcommand{\eeqa}{\end{eqnarray}}

\documentclass[twoside]{article}
\usepackage{graphicx,times,latexsym}

\begin{document}


\hfill{}

\vspace{2.0cm}

\begin{center}
{\large {\bf  Time of arrival in Classical and Quantum Mechanics}}

\vspace{1.2cm}

Juan Le\'on

\vspace{0.8cm}

Instituto de Matem\'aticas y F\'{\i}sica Fundamental, CSIC\\
Serrano 113-bis, E-28006 MADRID, Spain\\
leon@imaff.cfmac.csic.es\\[0.4cm]
\end{center}
\vspace{0.6cm}
\begin{center}\today \end{center}

\vspace{1cm}
\begin{abstract}
The time of arrival at an arbitrary position in configuration space can be given as a function of the phase space variables for the Liouville integrable systems of classical mechanics, but only for them. We review the Jacobi-Lie transformation that explicitly implements this function of positions and momenta.  We then discuss the recently developed quantum formalism for the time of arrival. We first analyze the case of free particles in one and three space dimensions. Then, we apply the quantum version of the Jacobi-Lie transformation to work out the time of arrival operator in the presence of interactions. We discuss the formalism and its interpretation. We finish by disclosing the presence (absence) of "instantaneous" tunneling for thin (thick) barriers.
\\[0.3cm]
\end{abstract}

\vspace{2cm} \vfill


Classical and Quantum Mechanics use the notion of Newtonian time, an universal parameter that rules the evolution of all the dynamical systems of the universe. Newtonian time is ``a priori" external to everything, physical systems and observers alike. However, in many instances there are true time-like properties in the physical systems under study. In general, the answers to questions like: How long will it take to ...?  or, When will it ...?  etc. come in the form of a time that genuinely depends of the very system. The crux of the matter is that of finding the time (the time elapsed,  or the instant in time) in which some property of the system will take a specific value, something that could be generically termed as ``the time of arrival at that value". In the next section we deal with the formulation of this question in classical dynamics. The much more involved case of translating a time parameter into an operator on the Hilbert space, as requires the quantum treatment, is worked out in sections 2 and 3 for free and interacting particles respectively. In section 4 we point out some eccentric properties of the time of arrival at places that are classically forbidden.

\section{Deriving time in phase space}
The treatment of time as a phase space variable is a time-honored procedure. The term extended phase space was coined for the approach in which, to the $n$ pairs $(q,p)$ of the phase space variables of mechanical systems with $n$ degrees of freedom, one adds the additional conjugate pair $(t,p_t)$, which requires the constraint $p_t+H(q,p)=0$ for consistency. It seems possible to dismantle this construction trading in the way the pair $(t,p_t)$ by  another canonical pair $(q,p)$. Naively, one would single out one of the phase space variables ($q_1$ for instance) and make it equal to some parametric value (i.e. $q_1=x$). Then, its canonical conjugate (the momentum $p_1$ in this example), would be fixed by the constraint, giving $p_1=\phi(x;q_2,p_2,\ldots,p_t)$. The phase space would now be given in terms of $\{(q_2,p_2),\ldots,(q_n,p_n),(t,p_t)\}$, with $x$ acting as an external evolution-like parameter. Hence, $t(x)$ or in words, ``the time of arrival at $x$" would be a legitimate question to ask. In spite of its apparent generality, is seldom possible to accomplish this program, not because of its very difficulty, but due to the non-fulfillment of some of the many conditions necessary for the existence of solutions. Here we will discuss the case of integrable systems for which there is a global construct for $t(x)$, that we will describe explicitly.

In the modern approach to classical dynamics (a standard reference is~\cite{Arnold}, a very readable text can be found in~\cite{McCauley}), a Hamiltonian system is called completely integrable (a l\`a Liouville) when it satisfies the conditions $a$ and $b$ below:
\begin{itemize}
\item[a.] There are $n$ compatible conservation laws\\
$\Phi_i(q_1,\ldots,q_n,p_1\ldots,p_n;t)=C_i$, $i=1,\ldots,n$, that
is:
\begin{itemize}
\item[a.1.]$\dot{\Phi}_i
=\{\Phi_i,H\}+\frac{\partial\Phi_i}{\partial t}=0,\;\forall\;
i=1,\ldots,n.$
\item[a.2.] $\{\Phi_i,\Phi_j\}=0,\; \forall \;i,j=1,\ldots,n.$
\end{itemize}
\item[b.] The conservation laws define $n$ isolating integrals that
can be written as:
\begin{itemize}
\item[b.1.] $\Phi_i=C_i \Rightarrow
p_i=\phi_i(q_1,\ldots,q_n,C_1,\ldots,C_n;t),\; \forall\;
i=1,\ldots,n.$
\item[b.2.] $\frac{\partial \phi_i}{\partial q_j}=
\frac{\partial \phi_j}{\partial q_i}\, \;\forall\; i,j=1,\ldots,n.$
\end{itemize}
\end{itemize}
In these conditions, the solution to the Hamilton equations is an
integrable flow, described by a system of holonomic coordinates
$(q(t),p(t))$ in phase space for each instant of time:
\begin{eqnarray}
q_i(t)&=q_i(q_0,p_0;t),&\;\;i=1,\ldots,n. \label{classic1}\\
p_i(t)&=p_i(q_0,p_0;t),&\;\;i=1,\ldots,n.\nonumber
\end{eqnarray}
In particular, given the set of initial conditions $(q_0,p_0)$, the system arrives at the point $(q(t),p(t))$ of the phase space in the path independent instant $t$. Conversely, these points define the corresponding times of arrival. In this case, time meets the requirements to qualify as a derived variable in phase space, whose explicit construction occupies the rest of this section.

For integrable flows there is a special choice of phase space coordinates that mathematically eliminates  the effects of interactions (because the new positions are ignorable coordinates). In other words, integrable systems are canonically equivalent to a set of translations (or of circular motions) at constant speed. It is customary to denote the variables that determine these translations as action-angle variables, a name which strictly is appropriate only for the case of periodic systems, where the (closed) flow lines are topologically equivalent to circles. For these integrable flows, there is a canonical transformation $W$ (the Jacobi-Lie transformation) to  free-like variables
\begin{equation}
\{q,p;H(q,p)\}\stackrel{W}{\longrightarrow}\{Q,P;H_0(P)\}
 \label{classic2}
\end{equation}
where $H(q,p)=H_0(P)$. The most useful form of this transformation is $W(q,P)$, that is, a function of the old positions and the new momenta, so that
\begin{equation}
Q_i=\frac{\partial W(q,P)}{\partial P_i},\;\; p_i=\frac{\partial
W(q,P)}{\partial q_i}  \; \;i=1, \ldots , n\label{classic3}
\end{equation}
The choice $H_0(P)=\sum_i \frac{P_i^2}{2m}$ relates the free coordinates $P_i(t)=P_i$ and $Q_i(t)=\frac{P_i}{m} t +Q_i$ of the translation flow to the positions and momenta $(q_i(t),p_i(t))$ of the actual flow generated by $H(q,p)=\sum_i \frac{p_i^2}{2 m}+V(q)$. In this work we shall only consider unbound systems with positive energy $H=H_0\geq 0$. For this reason we choose the conserved momenta $P_i$ for the constant variables  instead of the usual actions over a period $\oint p dq$ that are more apt for bounded motions. Notice however that the $P_i$ are different from the momenta appearing in perturbative calculations, even if both sets may coincide asymptotically or in some set of $R^n$. Coming back to our problem, the function $W$ would  be given explicitly as a complete integral of the following Hamilton-Jacobi equation:
\begin{equation}
H(q_i,\frac{\partial W(q,P)}{\partial
q_i})=\sum_{i=1}^n\frac{P_i^2}{2 m}\label{classic4}
\end{equation}
Due to the relations b.1 and b.2 above, it is permitted to write
$W$ as the path-independent integral:
 \beq
 W(q,P)=\sum_{i=1}^n \int_{q_0}^q dq_i
 \phi_i(q,C)\label{classic5}
 \eeq
where $q_0$ is a constant configuration space point, and the $C_i$ (that remain fixed during the integration) are functions of the $P_i$ whose determination is necessary to solve explicitly the problem. We are not concerned here with the search of specific solutions, but with the fact that integrability ensures their global existence. In fact, the equations (\ref{classic3}) can be written in the form
 \beq
 p_a(q,P)=\phi_a(q,C),\;\;
 Q_a(q,P)=\,\sum_{i=1}^n \int_{q_0}^q dq_i \frac{\partial
 p_i(q,P)}{\partial P_a},\;\; a=1,\ldots,n\label{classic6}
 \eeq
The first equation is simply the definition of the isolating
integrals (b.1). As a bonus, time can be given as a function of
phase space in two alternative ways: either in terms of the old
variables, or equally in terms of the new ones. Consider that a
particle initially at $(\mathbf{q,p})$ arrives at the position
${\mathbf{q}}(t)=\mathbf{x}$ in the instant $t({\mathbf{x}})=t$,
then:
\begin{equation}
t({\mathbf{x}})=\frac{m}{P_a}(X_a-Q_a)=\frac{m}{P_a}\sum_{i=1}^n
\int_{\mathbf{q}}^{\mathbf{x}}  dq_i \frac{\partial
 p_i({\mathbf{q,P}})}{\partial P_a},\;\; a=1,\ldots,n\label{classic7}
\end{equation}
where $X_a=\partial W({\mathbf{x,P}})/{\partial P_a}$ (obviously, $X_a=Q_a(t({\mathbf{x}}))$ by construction). Note that in (\ref{classic7}) there is no summation over the index $a$. In fact, integrability can be envisioned as the simultaneous existence of $n$ independent flows each of them contained in a different phase space plane. The requirement of integrability was noticed by Einstein~\cite{Einstein}, who analyzed its implications for the old Bohr-Sommerfeld quantization conditions, that he reformulated accordingly giving a new condition, that was criticized by Epstein~\cite{Epstein}. Integrability~\cite{Gutzwiller} allows $n$ different expressions to define the unique time of arrival. Only a pair $(Q_a,P_a)$ appears in each of them, and they all are equivalent. This holds even when there is no separable solution to the original Hamilton-Jacobi equation (\ref{classic4}) due to the presence of the potential $V(\mathbf{q})$ in the Hamiltonian. Only for some well known cases~\cite{Whittaker} the problem is separable in the original variables. Independently of this, notice that as $({\mathbf{Q}}(t),\mathbf{P})$ defines a straight line in the phase space, it is simple to lay one of the axes (the $n^{th}$ say) along it. This amounts to define $H_0({\mathbf{P}})=P_n$ which gives $P_n(t)=E$ and $Q_n(t)= t+Q_n$, while the other variables remain constant $Q_j(t)=Q_j,\, P_j(t)=P_j,\; j=1,\ldots,n-1$. With this choice, one can write: \beq t({\mathbf{x}})=\sum_{i=1}^n \int_q^x dq_i \frac{\partial p_i(q_1 \ldots q_n,P_1 \ldots P_{n-1},E)}{\partial E}\label{classic8} \eeq with the $p_i$'s given in (\ref{classic6}). This is the standard equation of time that appears in the literature. The rest of the relations would give the time independent geometric properties of the trajectories. Note that for central potentials only $p_r$ depends on $E$, so that (\ref{classic8}) reduces to $t(r_x)=\int_{r_0}^{r_x}dr (\partial p_r/\partial E)$.

We have focused the discussion of this section  on the dual definition of the time of arrival, that can be given in terms of the original phase space variables, or of the free translation variables. This duality is a foundation stone for the quantum method presented in this paper. We will obtain the time of arrival operator of interacting particles $\hat{t}(x)$ by applying a quantum version of the canonical transformation $W(q,P)$ to the well known operator for the time of arrival of free particles $\hat{t}_0(x)$. The properties of the latter have been extensively analyzed in the literature. For completeness, and to fix the notation, we present a summary of them in the next section.

\section{Time of arrival of free quantum particles}
In one space dimension Eq. (\ref{classic7}) gives the time of arrival at $x$ of a free particle initially at $(q,p)$ as a function of the phase space variables that depends on $x$ parametrically: $t_0(q,p;x)=m (x-q)/p$. In spite of its simplicity, this expression presents serious quantization difficulties~\cite{Pauli,Aharonov1,Paul,Kijowski,Galapon} whose solution we outline here~\cite{Kijowski,Tate,Juan,Giannitrapani1,Delgado}. First of all, it requires a decision about operator ordering, the simplest one being symmetrization:
\begin{equation}
  \hat{t}_0(\hat{q},\hat{p};x)=m (\frac{x}{\hat{p}}- \frac{1}{2}
\{\hat{q},\frac{1}{\hat{p}}\}_+)=-e^{-i{\hat{p}x}}\sqrt{\frac{m}
{\hat{p}}}\;\;{\hat{q}}\;\sqrt{\frac{m}{\hat{p}}}\;e^{i\hat{p}x}
\label{jj}
\end{equation}
Notice the proliferation of carets above. It is a reminder that we now deal with operators acting on the Hilbert space of the free particle states. From now on, we will drop the operator carets, simplifying the notation as much as possible, wherever this will not produce confusion between operators and c-number variables. The eigenstates $|t x s 0\rangle$ of this operator $t_0(x)$ in the momentum representation can be given as ($\hbar =1$)
\begin{equation}
\langle p|t x s 0\rangle=\theta(sp) \sqrt{\frac{|p|}{m}}\,\exp(i
\frac{p^2}{2m}t)\,\langle p | x \rangle \label{k155}
\end{equation}
where $t$ is the time eigenvalue, $x$ the arrival position, and
where we use $s=r$ for right-movers ($p>0$), and $s=l$ for
left-movers ($p<0$.) The label $0$ stands for free particle case.
Finally, the argument $sp$ of the step function that appears on the
rhs is $+p$ for $s=r$, and $-p$ for $s=l$, so that
 \beq
 \theta(rp)=\int_0^\infty dp |p\rangle \langle p|\;\;\mbox{and}\;\;
  \theta(lp)=\int_{-\infty}^0 dp |p\rangle \langle p|\label{free1}
  \eeq
The degeneracy of the energy with respect to the sign of the moment is explicitly shown by means of a label $s=r,l$ in the energy representation, where
\begin{equation}
\langle E s' 0|t x s 0\rangle = \delta_{s's}\,(\frac{2 E}{m})^{1/4}
e^{i E t}\, \langle E s 0 |x\rangle \label{k15}
\end{equation}

Summarizing, there is a representation for the time of arrival at
$x$ spanned by the eigenstates
\begin{equation}
|t x s 0\rangle=(\frac{2 H_0}{m})^{1/4} e^{i H_0 t} \Pi_{s0}
|x\rangle \label{k15a}
\end{equation}
where $\Pi_{s0}$ projects on the subspace of right-movers ($s=r$),
or of left-movers ($s=l$), i.e.
\begin{equation}
\Pi_{s0} =\int_0^{\infty} dE |E s 0\rangle \langle Es 0|=\theta(sp)
\label{k15b}
\end{equation}

These time eigenstates are not orthogonal. This gave rise in the past to serious doubts about their physical meaning. The origin of the problem can be traced back to the fact that (\ref{jj}) is not self-adjoint, that is, that $\langle\varphi|t_0(x) \psi\rangle \neq \langle t_0(x) \varphi| \psi\rangle$.  This was pointed out by Pauli~\cite{Pauli} long time ago and is due to the lower bound on the energy spectrum that prevents the applicability of the Stone theorem~\cite{Stone}. The problem emerges as soon as one attempts integration by parts in the energy representation.

Not being self-adjoint or orthogonal, this operator poses an interpretation problem that can be solved by considering it in terms the Positive Operator Valued Measures (POVM). This is a class of operators less restrictive than the traditional projector valued measures.  The POV measures only requires the hermiticity of $t_0(x)$ (i.e. $t_0(x)={(t_0(x))^*}^\top$) to assure the positivity of the measure. Now, instead of a Projector Valued spectral decomposition of the identity operator, one has the POV measure
\beqa
 P_0(\Pi(x);t_1,t_2)&=&\, \sum_s \int_1^2 dt \, |t x s 0\rangle
\,\langle t x s 0|\nonumber\\
&=&\,
 \sum_s \int_1^2 dt \,(\frac{2 H_0}{m})^{1/4}\, e^{i H_0 t}\,
 \Pi_{s0}\,\Pi(x)\,\Pi_{s0}\, e^{-i H_0 t}\,
 (\frac{2 H_0}{m})^{1/4}\label{k17}
\eeqa
 whose notation indicates the arrival interval and that the dependence on the arrival position comes through the projector $\Pi(x)=|x\rangle\, \langle x |$ on the position eigenstate. For the above measure $P_0(1,2)^2\neq P_0(1,2)$ because $|t x s 0\rangle \langle t x s 0|$ is not a projector, as the states are not orthogonal. However, the limit as $t\rightarrow \infty$ of $P_0(-t,+t)$ is the identity as can be checked explicitly. The time operator obtained is well suited for interpretation. This solution was introduced in~\cite{Giannitrapani1}, and extensively analyzed in refs. ~\cite{Giannitrapani2,Toller1}. It has been recently reviewed in ~\cite{Muga6} and criticized in~\cite{Kijowski2}.

In this formulation the time of arrival is given by the first
moment of the measure
 \beqa
 t_{0}(H_0, \Pi(x))\,&=&\, \sum_s \int_{-\infty}^{+\infty} dt \,t\; |t x s
0\rangle \,\langle t x s 0|\nonumber \\
& =&\,\int_{-\infty}^{+\infty} dt\, t\,
(\frac{2 H_0}{m})^{1/4}\, e^{i H_0 t}\, {\cal P}_0(x)\, e^{-i H_0
t}\, (\frac{2 H_0}{m})^{1/4}\label{free3}
 \eeqa
 where ${\cal P}_0(x)=\sum_s \Pi_{s0}\, \Pi(x)\, \Pi_{s0}$, which is not a projector. We now have the tools necessary for the physical interpretation of the formalism: Given an arbitrary state $\psi$ at $t=0$, its time of arrival at a position $x$ has to be, according to (\ref{free3}),
 \begin{equation}
\langle\psi|t_0(x)|\psi\rangle=\frac{1}{P_0(x)}\sum_s
\int_{-\infty}^{+\infty} dt\, t \, |\langle t x s 0|
\psi\rangle|^2,\label{k22}
\end{equation}
with the standard interpretation of $\sum_s|\langle t x s 0| \psi\rangle|^2$ like the (as yet unnormalized) probability density that the state $|\psi\rangle$ arrives at $x$ in the time $t$. The probability of arriving at $x$ at any time is then $P_0(x)=\int dt\; \sum_s|\langle t x s 0| \psi\rangle|^2$, giving a normalized probability density in times of arrival
\begin{equation}
P_0(t,x)=\frac{1}{P_0(x)}\;\sum_s|\langle t x s 0|
\psi\rangle|^2\label{k222}
\end{equation}
normalization that has been used in (\ref{k22}).

The above equations (\ref{k22},\ref{k222}) can be given  forms that are very useful for computation and that throw some light on the physical meaning of the different quantities involved. By using explicitly (\ref{k15a}), one gets
\begin{eqnarray}
P_0(x)&=&\sum_s \{\int dE\,(\frac{2E}{m})^{1/4} \langle x|E s
0\rangle \langle E s 0|\psi\rangle \}^*\,\nonumber \\
 &\times& \{\int dE'\, (\frac{2E'}{m})^{1/4} \langle x|E' s 0\rangle
\langle E' s 0|\psi\rangle \} \int dt \, e^{-i (E-E')t}\nonumber
\\ &=& 2 \pi \sum_s \int dE (\frac{2 E}{m})^{1/2} |\langle x|E s
0\rangle \langle E s 0|\psi\rangle|^2\label{k23}
\end{eqnarray}
The use of a similar procedure in (\ref{k22}) leads to $\;\;\;\langle \psi|t_0(x)|\psi\rangle$
\begin{eqnarray}
&=&-\frac{i \pi}{P_0(x)} \sum_s \int
dE (\frac{2 E}{m})^{1/2}\, \{\langle x|E s 0\rangle \langle E s
0|\psi\rangle \}^*\,
\stackrel{\longleftrightarrow}{\frac{\partial}{\partial
E}}\,\{\langle x|E s 0\rangle \langle E s 0|\psi\rangle
\}\label{k24}\\ &=& \frac{2 \pi}{P_0(x)} \sum_s \int dE (\frac{2
E}{m})^{1/2}\nonumber |\langle x|E s 0\rangle \langle E s
0|\psi\rangle |^2\, \frac{\partial}{\partial E}\{\arg\langle x|E s
0\rangle +\arg \langle E s 0|\psi\rangle\}\,
\end{eqnarray}
This expression is easy to understand. In fact it involves two ingredients: the plane wave amplitude $\langle x|E s 0\rangle=\sqrt{\frac{m}{2\pi p}} \exp(i s p x)$, along with the bracket $\langle E s 0|\psi\rangle =\sqrt{\frac{m}{p}} \tilde{\psi}(sp)$ where $\tilde{\psi}$ is the Fourier transform of the initial state in momentum space, and $p=\sqrt{2mE}$. This gives for the arrival amplitude
 \beq
 \langle t x s 0|\psi\rangle=\frac{1}{2\pi} \int_0^{\infty} dp\;
 \sqrt{\frac{p}{m}} e^{i(-E t+spx)} \tilde{\psi}(sp)\label{free4}
 \eeq
This is a free case so that the probability of ever arriving to $x$ has to be one. In fact
 \beq
P_0(x)=\sum_{s=r,l}\int_0^{\infty} dp\;|\tilde{\psi}(sp)|^2=
\int_{-\infty}^{+\infty} dp\;|\tilde{\psi}(p)|^2=1\label{free5}
\eeq where we used that in our notation $rp=+p$ and $lp=-p$. We
also have:
 \beq
\frac{\partial}{\partial E}\{\arg\langle x|E s 0\rangle +\arg
\langle E s 0|\psi\rangle\}=\frac{m}{p}  (sx-\frac{\partial
\arg(\tilde{\psi}(sp))}{\partial p})
 \label{free6}
 \eeq
There are initial wave packets centered around the values
$(q_0,p_0)$ for which $\tilde{\psi}(p)=|\tilde{\psi}(p)|\;\;
 \exp(-i p q_0)$
with the amplitude $\tilde{\psi}(p)$ peaked around $p_0$. Then,
$({\partial}/{\partial p}) \arg\tilde{\psi}(sp)=sq_0$, and the time
of arrival at $x$ reduces to \beq
 \langle \psi|t_0(x)|\psi\rangle=
 \sum_{s=r,l} \int_0^{\infty} dp\;
 |\tilde{\psi}(sp) |^2\; \frac{m  (x-q_0)}{sp}=\int_{-\infty}^{+\infty}
 dp\;|\tilde{\psi}(p) |^2\; \frac{m  (x-q_0)}{p}\label{free7}
  \eeq
which is the time of arrival of the classical free particle
averaged over its initial state.

The generalization to the case of three space dimensions~\cite{Juan} is not straightforward. The reason is that to begin with, there are three equivalent  equations  (\ref{classic7}) for the time of arrival. To be compatible, they have to satisfy the constraints
 \beq
 \mathbf{{\cal L}}=\, \mathbf{(q-x)}\wedge  \mathbf{p} =0 \label{three1}
 \eeq
where we drop the distinctions made in (\ref{classic7}) between upper and lower case letters, as they are the same objects for free particles. Classically, the constraints correspond to the fact that $\mathbf{x}$ has to be a point of the particle's trajectory, therefore the angular momentum can be written as $\mathbf{L}=\mathbf{x}\wedge \mathbf{p}$. In other words, the angular momentum with respect to $\mathbf{x}$, that is $ \mathbf{{\cal L}}$, has to vanish. We now show that the constraints (\ref{three1}) are first class. First of all, they are closed as their components ${\cal L}_a= \epsilon_{abc}\, (q-x)_b\, p_c$ satisfy the algebra of 3-D rotations, namely $\{ {\cal L}_a,{\cal L}_b\}=\epsilon_{abc}\,{\cal L}_c$. Then, the total Hamiltonian is $H_T=\frac{\mathbf{p}^2}{2\,m}+\mathbf{\lambda\cdot {\cal L}}$, where $\mathbf{\lambda}$ is a vector multiplier, so that $\{{\cal L}_a,H_T\}=\epsilon_{abc}\,\lambda_b\,{\cal L}_c$. Therefore, the constraints form a  first class system that depends parametrically on $\mathbf{x}$, one for each arrival position $\mathbf{x}$. Not all the  $\mathbf{x}$'s can be reached from an arbitrary set $(\mathbf{q},\mathbf{p})$ of phase space variables. Only those $\mathbf{x}$ that satisfy the constraints are positions where the particles with these dynamical variables can eventually be detected. A detector placed somewhere else will miss them.

The above translates into quantum mechanics as it is: Not all the states in the Hilbert space of free particle states ${\cal H}$ with Hamiltonian $H_0$ can be detected at a specific position $\mathbf{x}$. Only the subspace ${\cal H}_{\mathbf{x}}$ of the states that satisfy the constraints (\ref{three1}) (where now $\mathbf{q}$ and $\mathbf{p}$ are operators) qualify as the Hilbert space of detected states (at $\mathbf{x}$).  This subspace is spanned by the states $|\psi; \,\mathbf{x}\rangle \in {\cal H}$ of  the form $  |\psi;\, {\mathbf{x}} \rangle =\psi(H_0)\,  |\,{\mathbf{x}} \rangle$. Here, $\psi (H_0)$ is an arbitrary function of $H_0$, that may also depend on other parameters, $|\,{\mathbf{x}} \rangle$ is the  eigenstate ${\mathbf{q}}\,|\,{\mathbf{x}}\,\rangle= {\mathbf{x}}\,|\,{\mathbf{x}}\,\rangle$ of the arrival position. In particular, the detected subspace ${\cal H}_{\mathbf{x}}$ is obtained from ${\cal H}_{\mathbf{0}}$ by a translation of amount $\mathbf{x}$, as required by covariance.

The value of $t$ comes from the equation of motion in the subspace
orthogonal to the constraints, namely ${\mathbf{ p\cdot
x}}={\mathbf{ p}}\cdot \left(\frac{{\mathbf{ p}}}{m}\,
{t_0}({\mathbf{x}} ){+} { \mathbf{q}} \right)$, that in  spherical
coordinates where $|{\mathbf{ p}}
\rangle=|p,\theta_p,\phi_p\rangle$ and $q=i\frac{d}{dp}$ can be
written as
 $x= \frac{p}{m}\,t_0({\mathbf{ x}} )+q$.
 This can be readily inverted to give
\beq t_0({\mathbf{ x}})=-\frac{m}{p} \,e^{-i {\mathbf{ px}}}
p^{-1/2}\, q\, p^{1/2}\, e^{i {\mathbf{ px}}}\label{three2}\eeq
 Notice the characteristic powers of $p$ to the right and to the left of $q$. This operator ordering makes of $t_0$ a maximally symmetric operator with respect to the measure $d^3p$, making integration by parts a straightforward task. In $d$ space dimensions we would have $t_0\propto \frac{1}{p^{n+1}}\,(-i \frac{d}{dp})\,p^n$ with $n=(d-2)/2$~\cite{Juan}. The eigenfunctions of $t_0$ are given in the momentum representation by:
 \beq \langle {\mathbf{p}} | t;\, {\mathbf{x}},\,0 \rangle=
\sqrt{\frac{1}{4\pi m p}}\, e^{i E_{\mathbf{p}} t}
\langle{\mathbf{p}} |\,{\mathbf{x}} \rangle \label{three3} \eeq
 where $t\in\mathbf{R}$ is the time eigenvalue,
and $E_{\mathbf{p}}=p^2/2m$. One can define a time of arrival
representation given by
 \beq
| t;\, {\mathbf{x}}, \,0 \rangle=\frac{1}{\sqrt{4\pi m}}\,
 (\frac{1}{2\,m\,H_0})^{1/4} e^{iH_0\, t}\,|\,{\mathbf{x}}
 \rangle\label{three4} \eeq
These eigenstates are not orthogonal. They correspond to a POV
measure defined by the spectral decomposition
 \beq {\mathbf{1}}_{\mathbf{x}} =\int_{-\infty}^{+\infty} dt\, | t;\,
{\mathbf{x}}, \,0 \rangle \langle  t;\, {\mathbf{x}}, \,0\,|
\label{three5}\eeq
 It can be immediately seen that for any state
 $|\psi;{\mathbf{x}}\rangle \in {\cal H}_{\mathbf{x}}$, and for
 arbitrary momentum ${\mathbf{p}}$
 \beq
 \langle
 {\mathbf{p}}\,|\,{\mathbf{1}_{\mathbf{x}}}\,|\,\psi;{\mathbf{x}}\rangle=
 \langle  {\mathbf{p}}\,|\,\psi;{\mathbf{x}}\rangle\,\;\;
 \forall\,{\mathbf{p}}\in {\mathbf{R}}^3
 \label{three6}\eeq
 Therefore, the operator $\mathbf{1}_{\mathbf{x}}$ is a decomposition of the identity within the subspace of detected states ${\cal H}_{\mathbf{x}}$. The fact that ${\mathbf{1}}_{\mathbf{x}}<\mathbf{1}$, so that the decomposition is uncompleted, is the quantum version of the classical case where only a part of the incoming particles will (reach and) be detected at ${\mathbf{x}}$. From here it is clear that our formalism is finer than that provided by the so-called screen operators~\cite{Mielnik}, that would describe the arrival at a two dimensional plane put across the particle trajectories. In fact, these screen operators would correspond to a coarse graining of the present formalism, whose interpretation is analyzed in some detail in~\cite{Juan}.

 The time of arrival can be given through the first momentum of the
 POV measure (\ref{three5}):
 \beqa
 t_0({\mathbf{x}})&=&\,\int dt\,\,t\,\, | t;\,
{\mathbf{x}}, \,0 \rangle \langle  t;\, {\mathbf{x}}, \,0\,|
\nonumber \\ &=&\,
\frac{1}{{4\pi m}}\,\int dt\,\,t\,\,
 (\frac{1}{2\,m\,H_0})^{1/4} e^{iH_0\, t}\,|\,{\mathbf{x}}
 \rangle \langle \,{\mathbf{x}}\,|\,
  (\frac{1}{2\,m\,H_0})^{1/4} e^{-iH_0\, t}\label{three7}
  \eeqa
  whose similarity with the 1-D case (\ref{free3}) is
  evident, and can be used as a guide to get the average time of
  arrival an other quantities of interest that were worked on in
  one space dimension.

\section{The arrival of interacting particles}

In this section we want to determine the effect on the times of arrival of a position dependent interaction between the particle and the medium, that we describe by a potential energy $V(q)$. For instance, we want to consider the case of a barrier placed between the detector and the initial state. We would put a detector at $x$ (at the other side of the barrier), and prepare the initial state $|\psi\rangle$ of the particle at $t=0$ (at this side of the barrier). We would then record with a clock the time $t$ when the detector clicks. Repeating this procedure with identically prepared initial states, we would get the probability distribution $P(t,x)$ in times of arrival at $x$. This is the same procedure used for the free particle case, the  differences coming from the presence of the potential energy $V(q)$.

To find the quantum time of arrival we will use what we know from the classical case: There is a canonical transformation from the free ($H_0=\frac{P^2}{2m}$) translation variables $(Q,P)$ to the actual variables $(q,p)$ of the interacting situation where $H=\frac{p^2}{2m}+V(q)$. Time can be given equally in any of these two versions and we did already quantize the free version $t_0$ in the previous section. Now, in successive steps, we do the following~\cite{Leon}: We first construct the quantum canonical transformation $U$ that connects the free-particle states to the eigenstates of the complete Hamiltonian. This is the quantum version of the (inverse of the) Jacobi-Lie canonical transformation (\ref{classic2},\ref{classic5}). We will see later on that $U$ is given by the M\"oller wave operator. We will then apply  $U$ to $t_0$ to define the time of arrival $t$ in the presence of the interaction potential $V(q)$ in terms of $t_0$. We will work out the details of this transformation $t=U\, t_0\, U^\dagger$. Finally, we will also address some questions of interpretation of the resulting formalism.

Dirac introduced canonical transformations in quantum mechanics in a number of different places~\cite{Dirac} by means of unitary transformations $U$ ($U U^\dagger =U^\dagger U=1$ ). To fix the notation, we assume in what follows that the operators $q$ and $p$ are given in the coordinate representation of the Hilbert space $L^2(x)$ by $q=x$ and $p=-i \hbar \frac{\partial}{\partial x}$. If the operators $\bar{q}$ and $\bar{p}$ are the result of an arbitrary canonical transformation  applied to $ q$ and $p$, then there is a unitary transformation $U$ such that \beq
\bar q =U^\dagger \,q\, U,\;\;  \bar p  =U^\dagger\, p\, U\,
\Rightarrow \;[\bar q,\, \bar p ]=[q,\,p]=-i \hbar \label{k4} \eeq
One can also define implicitly the quantum canonical
transformations as is done in classical mechanics, a possibility
that has been thoroughly analyzed and developed. The main results
of the method are collected in~\cite{Moshinsky3}, which also
includes references to other relevant literature. The definition of $U$ is given implicitly by the two conditions
 \beq F(\bar q,\,\bar p)=F_0(q,\,p),\;\mbox{and} \;
 G(\bar q,\,\bar p)=G_0(q,\,p)\label{k44} \eeq
where $F,G,F_0$ and $G_0$ are functions of the operators shown explicitly as their arguments. They can not be chosen arbitrarily, the necessary and sufficient condition for the canonicity of the transformation being $ [F,\,G]=[F_0,\,G_0]$. The dependence of (\ref{k44}) on $U$ can be explicitly given by using (\ref{k4}) in it:
\beq
 U^\dagger\,  F(q,p)\,U  =\, F_0(q,p),\,\;\mbox{and} \;
U^\dagger\,G(q,p)\,U =\, G_0(q,p)  \label{k5} \eeq
 that comes from the straight application of (\ref{k4}) to the first members. In addition, $U$ is unitary so that the spectra of the original and transformed operators have to coincide. We now assume that $F$ and $F_0$ are self-adjoint operators whose eigenvalue problems are solved by the states $|f\,s\rangle$ and $|f\,s\,0\rangle$ (both corresponding to the same eigenvalue), that form orthogonal and complete bases of the Hilbert space satisfying
 \beq
F\,|f\,s\,\rangle=\lambda_f \,|f\,s\,\rangle,\;\;
F_0\,|f\,s\,0\rangle=\lambda_f \,|f\,s\,0\rangle\label{k7} \eeq
 We are accepting here the presence of  degeneracy indicated by the discrete index $s$, something that we will need later. Assuming now a continuous spectrum (the case we will be interested in), the operator $U$ that satisfies the first row of (\ref{k7}) is given by \beq U=\sum_s \int_{\sigma (\lambda)}  d\lambda_f |f\, s\rangle \langle  f\, s\,0| \label{knew} \eeq It is straightforward to verify that it is unitary. We can now give the definition of $ G$ in terms of $G_0$ using $U$, that is $G=U \,G_0\, U^\dagger$, which in full detail reads 
\begin{equation}
 G(q,p)=\sum_{ss'} \int_{\sigma({\lambda})} d\lambda_f
d\lambda_{f'}| f\, s\rangle \, \langle f\,
s\,0|G_0(q,p)|f'\,s'\,0\rangle \, \langle  f'\,s'| \label{k9}
\end{equation}
This is the main result of our procedure. The fact that we can
define an operator $G$, canonically conjugate to $F$, if we know
$G_0$ and $U$.

We will now apply this to the case where $F_0$ is the free Hamiltonian $H_0$, $F$ the complete Hamiltonian $H$ and $G_0$ the time of arrival $t_0$ of the free particle Eqs.(\ref{jj}), or (\ref{free3}). Then, we have $H_0=U^\dagger\, H\, U$ and $ \Pi_0(x)=U^\dagger\, \Pi(x)\, U$. Associated to the free particle there was the positive operator valued measure $P_0$ of Eq.(\ref{k17}). Accordingly, the POV measure $P$ of the interacting case will be given by ({\it cf} (\ref{k5}))
\begin{equation}
P(\Pi(x);t_1,t_2 )=U P_0 (\Pi_0(x);t_1,t_2) U^\dagger. \label{k9b}
\end{equation}
then, the time of arrival operator in the presence of interactions
(the $G$ of (\ref{k9})) is given by
\begin{equation}
t(H,\Pi (x))=U t_0 (H_0,\Pi_0(x)) U^\dagger. \label{k9a}
\end{equation}

We noticed above that the spectra of the original and transformed operators had to coincide. Now, $\sigma(H_0)=\mathbf{R^+}$ so that not all the Hamiltonians can be obtained from $H_0$ by this procedure. In general, some fixing will be required to make the spectra coincide. Here we will only consider well behaved potentials ($V(q)\geq 0\,\, \forall \, q\in\mathbf{R}$), vanishing appropriately at the spatial infinity. This ensures the required coincidence of the spectra, but introduces two solutions for $U$ due to the existence of two independent sets of eigenstates of $H$:
 \beq
U_{(\pm)}=\sum_s \int_0^{\infty} dE |E s (\pm)\rangle \langle E s
0|=\Omega_{(\pm)} \label{k10} \eeq
 these are the M\"oller operators connecting the  free particle states to the bound and scattering states. In presence the presence of bound states these operators would only be isometric, because the correspondence between eigenstates of $H$ and free states could not be one to one. In our case $V(q)\geq 0$, there is one free state for each scattering state and conversely. Thence, the M\"oller operators are unitary. In this case, the intertwining relations $H \Omega_{(\pm)}=\Omega_{(\pm)}H_0$ can be put in the more desirable form $H =\Omega_{(\pm)} H_0\Omega^\dagger_{(\pm)}$. We will also follow the standard sign conventions, choosing $\Omega_{(+)}$  in (\ref{k10}) that, when $E=\lim_{\epsilon\rightarrow 0^+}(E+i\epsilon)$, gives signal propagation forward in time. The results that would be obtained with $\Omega_{(-)}$ would correspond to the time reversal of this situation. If $\tau$ is the time reversal operator, then $P_{(-)}(\Pi(x);t_1,t_2 )=\tau \; P_{(+)}(\Pi(x);-t_2,-t_1 )\; \tau^\dagger$. For notation simplicity, we will omit these labels $(\pm )$ wherever possible.

The parameter $x$ that appears in (\ref{k9}) and (\ref{k9a}) is the actual detection position in the interacting case, the place whose time of arrival at we want to know. Therefore, the arguments of $t$ in (\ref{k9a}) have to be $\Pi(x)= |x\rangle \langle x|$ and $H$. Hence, the argument of $t_0$ will be  an object $\Pi_0(x)=\Omega^\dagger \Pi(x) \Omega$ which collects all the states of the free particle that add up to produce the actual position eigenstate $|x\rangle$  by the canonical transformation. Much of the difference between the classic and quantum cases is hidden here. In particular, the quantum capability to undergo classically forbidden jumps in phase space has much to do with the fact that $\Pi(x)$ and $\Pi_0(x)$ can not be position projection operators simultaneously.

We have now at hand all the tools necessary to answer the questions about the time of arrival of interacting particles. Given a particle that was initially (at $t=0$) prepared in the state $|\psi\rangle$, we can compute the predictions for the average time of arrival $\langle \psi |t(x)|\psi\rangle$, the probability distribution in times of arrival $P(t,x)$ and the probability of ever arriving at $x$, $P(x)$. Instead of writing more equations, we refer the reader to Eqs. (\ref{free3},\ref{k22}), (\ref{k17},\ref{k222}) and (\ref{k23}). By simply erasing the label $0$ from them, one gets the correct expressions for the interacting case, with the caveat that -to be of practical use- they require the knowledge of the scattering states and M\"oller operator. It is worth to recall here that the expression (\ref{k24}) for the average time remains valid after dropping the $0$'s. So, $\langle\psi|t(x)|\psi\rangle$ is still the sum of two independent pieces, one containing $(\partial/\partial E)\arg\langle E\,s\,| \psi\rangle$ that only depends of the initial state, the other that contains $(\partial/\partial E)\arg\langle x\,| E\,s\,\rangle$ and only depends of the position of arrival.

We now consider the case where there is a finite potential energy starting at the origin ($V(q)=0,\, \forall q\leq 0$), which is  so smooth that the quasi-classical approximation is valid. Then for $E>V(x)$ the exponentially small reflection amplitude can be neglected, giving the scattering states
\begin{equation}
\langle x|E\, r\, \rangle\approx\theta(-x)\,\sqrt{\frac{m}{2\pi
p}}\, e^{ipx}\,+\, \theta(x)\,\sqrt{\frac{m}{2\pi p(x)}}\,
e^{i\int_0^x dq\, p(q)}\label{n3}
\end{equation}
with $p(q)=\sqrt{2m(E-V(q))}$, that are normalized to an incoming right-moving particle by time unit. We now consider the physically interesting case where the initial wave packet is normalized to 1, (i.e. that $\int dp |\tilde{\psi}(p)|^2=1$ with $\tilde{\psi}(p)=\langle p|\psi\rangle$), also, that it is localized around a position $q_0$ well to the left of the origin, and that it has a mean momentum $p_0\gg V(x)$. Then, to this order the probability of ever arriving at $x$ (c.f. (\ref{k23})) gives
\begin{equation}
P(x)\approx \theta(-x)+\theta(x)\, P_>(x),\;\mbox{where}\;
P_>(x)=\int_0^\infty dp
\frac{p}{p(x)}\;|\tilde{\psi}(p)|^2\label{n4}
\end{equation}
so that $\frac{p}{p(x)}\;|\tilde{\psi}(p)|^2$ is the (unnormalized)
probability of arrival at the point $x$ with momentum $p(x)$, as corresponds to the quasiclassical case. Notice that to the left of the origin the result is the same that in the free case. This comes about because the approximation neglects  reflection, thus missing at $q<0$ any information about a the existence of a finite $V$ at $q\geq 0$. For the time probability distribution one gets
 \beqa
  P(t,x)&\approx&\,\frac{\theta(-x)}{2\pi}\left| \int_0^\infty dp\,
e^{-iEt}\;\tilde{\psi}(p)\right|^2 \nonumber \\
&+&\,\frac{\theta(x)}{2\pi
P_>(x)}\left|\int_0^\infty dE\;
\sqrt{\frac{m}{p(x)}}\;\tilde{\psi}(p)\;  e^{-i(Et-\int_0^x dq \
p(q))}\;\right|^2 \label{n5} \eeqa
 which, not surprisingly, is the same as that of free particles for
 $x<0$. Finally,
\beqa
 \langle\psi|{\bf
t}_x|\psi\rangle&\approx & \theta(-x) \int_0^\infty dp\;
|\tilde{\psi}(p)|^2 \;\frac{m}{p}\{x-q_0\}\,
\nonumber \\ &+&
\,\frac{\theta(x)}{P_>(x)}\int_0^\infty dp\,
\frac{p}{p(x)}\;|\tilde{\psi}(p)|^2\; \{-\frac{mq_0}{p}+m \int_0^x
\frac{ dq}{p(q)}\}\label{n6}
 \eeqa
 Therefore,  we recover the time of arrival of the free particles for negative $x$. On the other hand, for $x>0$ we get the classical time of arrival at $x$  for initial conditions $(q_0,p)$, $\int_{q_0}^x \,(m/p(q))\, dq$, weighted by the probability  of these conditions.

\section{Advanced or delayed arrival?}
What is the effect of putting a quantum barrier in the path of the arriving particle? Hartman~\cite{Hartman} studied this question a long time ago, reaching the conclusion that tunneled particles should appear instantaneously on the other side of the barrier. Our formalism supports this result, but only for thin enough barriers. 

The time of arrival at a point $x$ in the presence of a barrier will be given through a probability amplitude
\begin{equation}
\langle t x s|\psi\rangle = \int dE (\frac{2E}{m})^{1/4} e^{-i
Et}\, \langle x|Es(+)\rangle \langle Es(+)|\psi\rangle \label{psi2}
\end{equation}
In the case where $x$ is at the right of the barrier, the
amplitude can be approximately given by~\cite{Leon}
\begin{equation}
\langle t x s|\psi\rangle \approx
\frac{\delta_{sr}}{\sqrt{2\pi}}\int dE (\frac{m}{2E})^{1/4} e^{-i
(Et-px)} T(p) \tilde{\psi}(p)\label{psi3}
\end{equation}
where $T(p)$ is the transmission amplitude for momentum $p$.
Now, the total probability
of eventually arriving at $x$ in any time $t$ is
$P(x)\approx \int dp\; |T(p)\;
\tilde{\psi}(p)|^2$
that is independent of $x$ in cases like this, where $x$ is beyond the range of the potential. After a straightforward calculation we get for the average time of arrival at the other side of the barrier the corresponding version of (\ref{k24})
\begin{equation}
\langle\psi|{\bf t}_x|\psi\rangle\approx  \frac{1}{ P(x)}
\int_0^{\infty} dp\; |T(p)\, \tilde{\psi}(p)|^2 \frac{m}{p}
\{x-q_0+\frac{d\arg(T(p))}{dp}\} \label{psi55}
\end{equation}
 It is the value of the Wigner 
time~\cite{Wigner} averaged over the transmitted state.

Consider a simple square barrier of height $V$ and width $a$. 
The transmission coefficient is in this case:
\begin{equation}
T(p)= e^{-i p a}  \;\left(1-i \frac{(p^2 +p'^2 )}
{2 p p'} \tan p'a\right)^{-1}\; \sec p' a \label{t1}
 \end{equation}
 where $p'=\sqrt{p^2-p_V^2}$, that is imaginary for $p$ below $p_V$. Notice the contribution $-pa$ to $\arg(T(p))$. This will subtract a term $a$ to the path length $x-q_0$ that appears in (\ref{psi55}). The barrier has effective zero width or, in other words, it is traversed instantaneously. This is the Hartman effect for barriers. To be precise, the effect is not complete, it is compensated by the other dependences in $p'a$ present in the phase of $T(p)$. In fact, it disappears for low barriers $(p_V/p)\rightarrow 0$, where all the $a$ dependences of the phase cancel out, as was to be expected because the barrier effectively vanishes in this limit. In the opposite case of high barriers  $(p/p_V)\rightarrow 0$ the effect saturates and there is a  decrease $ -\frac{m a}{p}$ in the time of arrival of transmitted plane waves, that emerge almost instantaneously at the other side of the barrier.

The averaging of the Wigner time over the transmitted state, present in (\ref{psi55}) as a consequence of the formalism,  has dramatic effects, because it effectively forbids the transmission of the  wave components with low momenta. In fact, it produces the exponential suppression (by a factor $\exp(-2|p'|a)$) of the tunneled components. Therefore,  only the  components with momentum above $p_V$ have a chance of surmounting thick barriers, being  finally transmitted. But these components are delayed by the barrier (for them $(\frac{d\arg(T(p))}{dp})>0$), whose overall effect transmutes from advancement into retardation~\cite{Leon} at a definite predictable thickness that depends on the barrier height and also on the  properties of the incoming state.

\section*{Acknowledgements}
Parts of this work are based on a talk given at the I Cracow-Clausthal Workshop on Tunneling Effect and Other Fundamental Problems in Quantum Physics. The author is indebted to Jacek Jakiel and Edward Kapuscik for the warm hospitality extended to him during the Workshop. This work has been partially supported by the DGESIC under contract PB97-1256.

\end{document}